\documentstyle[12pt]{article}
\begin{document}

\title{MODERNAS TEORIAS SOBRE EL TIEMPO DISCRETO}

\author{Miguel Lorente \\
Universidad de Oviedo}
\date{}

\maketitle

\section{Introducci\'on}

En las explicaciones f\'{\i}sicas del Universo el tiempo aparece como una magnitud
fundamental que es imprescindible para describir los procesos naturales.
Cl\'asicamente el tiempo era una entidad absoluta, independiente de las cosas y que
de alguna manera acompa\~naba su devenir.

La teor\'{\i}a de la relatividad introdujo un car\'acter relativo en el tiempo
dependiente del observador, aunque esto no supon\'{\i}a un rechazo de la existencia
de un tiempo independiente de las cosas. Modernamente se han propuesto teor\'{\i}as
alternativas del tiempo como un concepto derivado de las relaciones entre las cosas
de modo que niega toda entidad al tiempo que no sea la misma realidad de las cosas.

Aunque estas teor\'{\i}as puedan parecer nuevas, hay que remontarse hasta Leibniz (y
a\'un m\'as lejos) ya que \'este propuso y defendi\'o ac\'erri\-ma\-men\-te contra
los disc\'{\i}pulos de Newton una teor\'{\i}a relacional del tiempo [1]. Obviamente
estas teor\'{\i}as del tiempo est\'an unidas a las del espacio por lo que es
pr\'acticamente imposible hablar de unas sin mencionar a las otras [2].

Siguiendo un orden epistemol\'ogico nos parece m\'as conveniente empezar por las
teor\'{\i}as f\'{\i}sicas que describen el tiempo como una magnitud discreta, cuya
finalidad es puramente pragm\'atica: resolver un conjunto de problemas
f\'{\i}sico-matem\'aticos sin preocuparse de su interpretaci\'on filos\'ofica.

Otro bloque de teor\'{\i}as intentan dar un contenido f\'{\i}sico a los modelos
discretos, buscando una estructura relacional de las cosas como substrato material a
las propiedades espacio-temporales.

Por \'ultimo se encuentran las posturas filos\'oficas que tratan de dar una
fundamentaci\'on ontol\'ogica a las teor\'{\i}as relacionales del espacio-tiempo.

\section{Modelos f\'{\i}sicos con tiempo discreto}

El uso de ret\'{\i}culos espaciales ha sido muy utilizado en F\'{\i}sica para poder
contar las part\'{\i}culas en cada celdilla y describir as\'{\i} las propiedades
estad\'{\i}sticas de un sistema con un n\'umero muy elevado de elementos. Hoy
d\'{\i}a se est\'a poniendo de moda el introducir valores discretos de las
coordenadas espacio-temporales en las ecuaciones que rigen la evoluci\'on de las
funciones de onda que representan part\'{\i}culas elementales en estado libre o en
interacci\'on. Estos modelos tienen un inter\'es matem\'atico: resolver por
m\'etodos num\'ericos ecuaciones diferenciales que no tienen una soluci\'on
anal\'{\i}tica exacta y por otra parte evitar los valores infinitos que aparecen en
los desarrollos perturbativos de dichas ecuaciones. La literatura reciente en esta
direcci\'on es muy amplia.

Para fen\'omenos locales donde se usa la relatividad especial, encontramos las
teor\'{\i}as gauge en ret\'{\i}culos rectangulares espacio-temporales que han sido
muy \'utiles para calcular factores de forma y masas de part\'{\i}culas elementales.

El trabajo pionero en esta l\'{\i}nea fue escrito por K. Wilson en 1974 que
intentaba explicar el confinamiento de los quarks en los hadrones y fue seguido por
otros f\'{\i}sicos en la descripci\'on de las interacciones fuertes y
electrod\'ebiles.

Al extender estos modelos a las interacciones gravitacionales, el ret\'{\i}culo
c\'ubico resulta insuficiente ya que es necesario adaptar al modelo las propiedades
geom\'e\-tri\-cas de los espacios curvos riemannianos usados en la relatividad general.
Uno de los primeros en proponer una teor\'{\i}a de espacio-tiempo discreto fue
Wheeler [3] que introdujo la discretizaci\'on del espacio-tiempo para abordar el
problema de la gravitaci\'on c\'uantica. Entre sus m\'as inmediatos seguidores se
encuentran Ponzano y Regge [4] que construyen un modelo de espacio-tiempo discreto,
donde una red de tri\'angulos adyacentes da lugar a superficies de curvatura
arbitraria. Recientemente han proliferado los autores que han seguido el  c\'alculo
de Regge [5].

La t\'ecnica general de estos modelos consiste en aproximar una superficie
Riemanniana por triangulaciones hechas de figuras simpliciales de lados iguales (un
simplicial es un conjunto de puntos donde cada uno de ellos est\'a relacionado por
aristas con todos los dem\'as). Cada capa formada de simpliciales est\'a enlazada a
otra capa pr\'oxima de la misma estructura de modo que hay una conexi\'on uno a uno
entre los puntos correspondientes de cada capa. La aplicaci\'on sucesiva de las
diferentes capas define una l\'{\i}nea del universo (en la terminolog\'{\i}a
relativista) cuyo par\'ametro para enumerar las diferentes capas toma valores
discretos y se puede identificar con el tiempo [6].

\section{Teor\'{\i}as relacionales del espacio-tiempo}

Estas teor\'{\i}as dan un paso m\'as en la explicaci\'on del Universo. Intentan
encontrar un modelo donde el espacio-tiempo es un concepto derivado de las
propiedades de las cosas y de las relaciones entre ellas. Tambi\'en aqu\'{\i}
Wheeler [7] fue precursor con su Pregeometr\'{\i}a basada en un conjunto de
constantes fundamentales con las que se describen las interacciones entre los 
``ladrillos'' o entidades b\'asicas del Universo. En la misma direcci\'on Marlow [8]
ha desarrollado una teor\'{\i}a axiom\'atica de la relatividad general cu\'antica basada en 
el c\'alculo de proposiciones donde el tiempo toma valores discretos.

Un autor que recogi\'o las sugerencias de Wheeler fue Penrose [9] que utiliz\'o las
entidades b\'asicas como unidades provistas de un valor de spin, que interaccionan
entre s\'{\i} siguiendo la ley de suma de dos momentos angulares. El resultado es
una red que no requiere un substrato espacial, porque ella es el mismo substrato. El
tiempo es un par\'ametro que enumera las diferentes interacciones que se producen
sucesivamente.

El programa de Penrose se prolonga en redes muy complejas que dan lugar a objetos
matem\'aticos quebautiz\'o con el nombre de ``twistors'' y que \'el
mismo utiliza en el formalismo de la teor\'{\i}a de la gravitaci\'on. Un enfoque
paralelo se puede encontrar en las ideas de D. Finkelstein [10]. Recientemente 
Penrose ha manifestado su intenci\'on de no renunciar a su modelo de espacio-tiempo
basado en una red de spines: ``Tengo todav\'{\i}a aspiraciones en mis ideas que he 
desarrollado hace varios a\~nos con la teor\'{\i}a de la red de spines (1971, Quantum 
theory and Beyond; 1972, Magic without magic). Los experimentos ideales del tipo de 
Bohm, Einstein, Podolsky y Rosen han jugado un papel muy importante en esa teor\'{\i}a, 
y la idea fue construir los conceptos de espacio y tiempo como una estructura l\'{\i}mite 
impl\'{\i}cita cuando el n\'umero de part\'{\i}culas se hace muy grande. Sin 
embargo, ni la teor\'{\i}a de los `twistors' ni la teor\'{\i}a de las `redes de 
spin' tienen entre sus ingredientes una asimetr\'{\i}a temporal. Por eso me resulta 
evidente que es necesario una idea esencialmente nueva'' [11].

Garc\'{\i}a Sucre y Bunge tambi\'en han introducido una teor\'{\i}a relacional del
espacio tiempo [12]. Las entidades fundamentales son todas las cosas del Universo
cuyas relaciones son descritas con un formalismo basado en la teor\'{\i}a de
conjuntos. Las unidades fundamentales o prepart\'{\i}culas se describen por
elementos de un conjunto finito. El papel crucial que juega el tiempo se representa
por la sucesi\'on de subconjuntos enlazados por la relaci\'on l\'ogica de la
inclusi\'on que implica un orden entre los mismos. El espacio no es m\'as que la
suma de todas las cadenas de subconjuntos ordenados por la inclusi\'on y que pueden
tomar todas las configuraciones posibles. Si escogemos unas determinadas l\'{\i}neas
entre todas las cadenas posibles habremos definido un sistema de referencia
determinado. El concepto de tiempo y de espacio emana de una manera natural de un
determinado sistema referencial.

Klapunosky y Weinstein [13] han propuesto recientemente una teor\'{\i}a de campos
cuantificados en los que los valores de las coordenadas espacio-temporales son
n\'umeros enteros y no representan ninguna referencia al espacio tiempo, sino unos
par\'ametros para distinguir los valores del campo. Las interacciones entre los
campos est\'an producidas por acoplamientos entre los campos fundamentales, de
manera que \'estos son la \'unica realidad subyacente y las relaciones producidas
por las conexiones entre ellos da lugar a un ret\'{\i}culo de estructura simplicial
que se nos presenta a los sentidos como una ``ilusi\'on'' que llamamos
espacio-tiempo.

El autor de este trabajo tambi\'en ha propuesto una teor\'{\i}a relacional del
espacio-tiempo [14] con el fin de justificar de una manera axiom\'atica los
fundamentos de la geometr\'{\i}a, m\'as all\'a todav\'{\i}a de los postulados
formulados por Hilbert. Si\-guien\-do el esp\'{\i}ritu de este matem\'atico, seg\'un el
cual, se deben considerar los puntos, l\'{\i}neas y superficies como sillas, mesas y
jarros de cerveza, (es decir, sin referencia a una intuici\'on espacial) se postula un ret\'{\i}culo n-dimensional c\'ubico donde cada
punto est\'a relacionado con 2n-puntos diferentes y solamente con \'estos, de cuyo
\'unico postulado se deducen l\'ogicamente todos los axiomas de Hilbert en su libro
Fundamentos de la Geometr\'{\i}a.

\section{Concepciones ontol\'ogicas subyacentes a las teo\-r\'{\i}as relacionales
del espacio-tiempo}

Las teor\'{\i}as relacionales mencionadas anteriormente se pueden analizar a un
nivel puramente l\'ogico (el tiempo que percibimos, se puede interpretar como la
impresi\'on sensible que nos produce la sucesi\'on temporal de relaciones entre los
objetos f\'{\i}sicos). Pero tambi\'en se puede preguntar sobre el substrato
ontol\'ogico de estas relaciones, que no suponga ninguna entidad fuera de las cosas
mismas.

Se puede citar a Leibniz entre los que han propuesto una explicaci\'on filos\'ofica
de la teor\'{\i}a relacional del espacio-tiempo. En su obra {\it Initia rerum
mathematicarum metaphysica}, defiende que el tiempo es el orden de las cosas
existentes que no son simult\'aneas, mientras que el espacio es el orden de las
cosas que coexisten o el orden de las cosas existentes que son simult\'aneas. El
fundamento del orden temporal es la conexi\'on causal. Cuando una cosa es el
principio de otra, aquella se dice anterior y \'esta posterior. Esta idea la vuelve
a repetir en su {\it Monadolog\'{\i}a} y en numerosas cartas.

Max Jammer [15] indica que Leibniz se inspir\'o para su  Monadolog\'{\i}a en la {\it
Gu\'{\i}a de perplejos} de Maim\'onides y recientemente Pannenberg [16] recuerda la
influencia de los fil\'osofos \'arabes en la teor\'{\i}a atomista del tiempo de
Leibniz. Seg\'un estos fil\'osofos la creencia en la creaci\'on implicaba que no
exist\'{\i}a nada antes de la creaci\'on y que los primeros seres existentes
ser\'{\i}an unos \'atomos o part\'{\i}culas indivisibles. El tiempo comenz\'o
tambi\'en en ese instante pero no como distinto de la materia sino concomitante con
ella. ``El tiempo, dice Maim\'onides [17], consta de instantes, a saber, que hay
mucha unidad de temporaneidad, los cuales por su ef\'{\i}mera duraci\'on, excluyen
la divisi\'on''. Hablar de \'atomos de materia es hablar de \'atomos de tiempo y por
consiguiente, a\~nade Maim\'onides ``el tiempo se inserta en `instantes' que no 
admiten divisi\'on''.

Dos autores contempor\'aneos que se pueden adscribir a una concepci\'on atomista del
tiempo son Whitehead y Weizs\"{a}ecker. Whitehead [18], el colaborador m\'as
estrecho de Russel, desarrolla en su edad madura una filosof\'{\i}a del Cosmos,
donde la \'ultima verdad es el atomismo metaf\'{\i}sico. Las entidades actuales
---denominadas tambi\'en ocasiones actuales--- son las \'ultimas cosas de que est\'a
compuesto el mundo. Las entidades actuales se relacionan entre s\'{\i} por nexos
extr\'{\i}nsecos e intr\'{\i}nsecos para formar estructuras m\'as complejas a
trav\'es de las prehensiones o sentires f\'{\i}sicos. Las entidades actuales
producen su tiempo y su espacio. La entidad actual es indivisible. La regi\'on que
ocupa la entidad actual es divisible s\'olo mentalmente. El tiempo y el espacio son
una abstracci\'on a partir de las actualidades y las relaciones entre ellos.

Para Weizs\"{a}ecker [19] los conceptos de espacio y tiempo son una consecuencia de
las relaciones entre las entidades m\'as fundamentales del Universo: los procesos
que el observador percibe como una simple alternativa (experimento si-no). Los
procesos constituyen un entramado de simples alternativas (``urs'') y el tiempo es
un par\'ametro que diversifica la realidad presente y futura de estos procesos.

Para Weizs\"{a}ecker la estructura actual del Universo est\'a compuesta de un
n\'umero finito de procesos elementales pero el n\'umero de posibilidades de
interacciones entre estos entes elementales es infinita, de donde se sigue el
car\'acter discreto para la descripci\'on de los entes actuales y continuo para las
leyes de evoluci\'on de estos procesos. Durante varios a\~nos Weizs\"{a}ecker ha
organizado unos Encuentros para trabajar en la unificaci\'on de la Mec\'anica
C\'uantica y la Relatividad de modo que los postulados de la \'ultima se derivan de
los postulados de la primera.

Como resumen de esta exposici\'on podemos decir que la hip\'otesis de un tiempo
discreto, como consecuencia de un concepto relacional del espacio-tiempo, se ha
desarrollado recientemente por numerosos autores en sus aspectos epistemol\'ogicos y
ontol\'ogicos, como alternativa a la concepci\'on absolutista, y que se corrobora
con la extensa bibliograf\'{\i}a.

Las consecuencias f\'{\i}sicas de esta hip\'otesis est\'an todav\'{\i}a muy lejos de
ser comprobadas experimentalmente aunque han progresado los modelos matem\'aticos
que permitir\'{\i}an hacer una predicci\'on detectable, por lo menos indirectamente,
de la hip\'otesis, y ciertamente mucho m\'as cercana a los hechos que las primitivas
especulaciones defendidas por Leibniz.

\vskip 1cm

\noindent {\large \bf  Referencias}

\medskip

\begin{enumerate}

\item[1] {\sc J. Earman}, {\it World enough and Space-time: Absolute versus
Relational Theories of Space and Time}, MIT Press, Cambridge 1989.

\item[2] {\sc M. Lorente}, ``Modernas teor\'{\i}as sobre la estructura del
espacio-tiempo'' en {\it Actas de la Reuni\'on Matem\'atica en honor de A. Dou},
Ed. Universidad Complutense, Madrid 1989, pp. 353--363.

\item[3] {\sc A. Wheeler}, {\it Geometrodynamics}, Academic Press, N.Y. 1962.

\item[4] {\sc	G. Ponzano, R. Regge}, en {\it Spectroscopy and Group Theoretical
Methods in Physics }(ed. F. Bloch), North Holland, Amsterdam 1968.

\item[5] {\sc	N.J. Lafave}, {\it A Step Toward Pregeometry I: Ponzano-Regge Spin
Networks and the Origin of Space-time Structure in Four Dimensions} (preprint),
Houston, Texas 1993.

\item[6] {\sc	Y. Shamir}, {\it Dynamical-Space Regular-Time Lattice and Induced
Gravity}, (preprint) Weizmann Institute of Science, Israel 1994.

\item[7] {\sc	A. Wheeler}, {\it Quantum Theory and Gravitation }(ed. A.R. Marlow)
Academic Press, N.Y. 1980.

\item[8] {\sc	A.R. Marlow}, ``An axiomatic general relativity quantum theory'' 
Cfr. [7] p. 35.

\item[9] {\sc	R. Penrose}, ``Angular Momentum: an Approach to Combinatorial
Space-Time'' en {\it Quantum Theory and Beyond }(T. Bastin, ed.) Cambridge U. Press,
1971.

\item[ ] {\sc R. Penrose}, ``On the nature of quantum geometry'' en {\it Magic 
without magic} (J.R. Klauder ed.) Freeman 1972.

\item[ ] {\sc R. Penrose}, ``On the origin of twistor theory'' en {\it Gravitation
and Geometry} (ed. W. Rindler and A. Trautman) Bibliopolis, Naples 1986.

\item[10] {\sc	D. Finkelstein, E. Rodr\'{\i}guez}, ``Quantum Time-Space and Gravity''
en {\it Quantum Concepts in Space and Time }(ed. R. Penrose, C.J. Isham) Clarendon
Press, Oxford 1986.

\item[11] {\sc R. Penrose}, ``Newton, quantum theory and reality'' en {\it Three 
hundred years of gravitation} (S.W. Hawking and W. Israel, ed.) Cambridge U. Press 
1987.

\item[12] {\sc G. Sucre}, ``Quantum Statistics in a Simple Model of Space-Time'',
{\it Int. J. of Theor. Phys.} 24, 441--445 (1985).

\item[ ] {\sc M. Bunge}, ``Una teor\'{\i}a relacional del espacio f\'{\i}sico, en
{\it Controversias en F\'{\i}sica}, Tecnos Madrid 1983.

\item[13] {\sc	V. Kaplunosky, M. Weinstein}, ``Space-time, Arena or Illusion'', {\it
Phys. Rev.} D 31, 1879--1898 (1985).

\item[14] {\sc	M. Lorente}, ``Quantum Processes and the Foundation of Relational
Theories of Space and Time''. {\it Encuentros Relativistas Espa\~{n}oles 1993
}(ser\'a publicado en Ed. Lumi\'ere, Par\'{\i}s 1994).

\item[15] {\sc	M. Jammer}, {\it Concepts of Space, The History of theories of 
Space in Physics}, Harvard U. Press,1969, p. 64.
Cambridge 1969.

\item[16] {\sc	Pannenberg}, {\it Systematische Theologie}, G\"ottingen 1991.

\item[17] {\sc	M. Maimonides}, {\it Gu\'{\i}a de Perplejos }(edici\'on preparada por
D. Gonz\'alez Maeso), Ed. Nacional Madrid 1983, p. 213.

\item[18] {\sc	A. Whitehead}, {\it The Concept of Nature}, Cambridge U. Press, 1920.
{\it Science and the Modern World}, McMillan 1925. {\it Process and Reality},
McMillan 1929.

\item[19] {\sc	K.F. Weizs\"{a}ecker}, {\it Die Einheit der Natur} (Hauser 1971) {\it
Quantum Theory and the Structure of Space and Time } 6 vol. (Hauser 1986) {\it Aufbau
der Physik}, Hauser 1985.

\end{enumerate}

\newpage

\noindent {\large \bf  Coloquio  a la comunicaci\'on de M. Lorente}

\medskip

En el coloquio subsiguiente, Alberto Dou manifest\'o su simpat\'{\i}a por estas teor\'{\i}as modernas del
tiempo discreto, porque eran coherentes con la progresiva cuantificaci\'on que la ciencia ha ido
ampliando en su descripci\'on de la naturaleza. Primero el atomismo de la materia fue introduciendo
unidades naturales en la composici\'on de los cuerpos, que se ha ido completando con las teor\'{\i}as de
part\'{\i}culas elementales, \'ultimos elementos invisibles de la materia. Por otro lado, la mec\'anica
cu\'antica ha introducido valores discretos naturales en ciertas magnitudes f\'{\i}sicas, como la carga,
el momento angular, la acci\'on. A. Dou describi\'o a continuaci\'on un modelo de estructura de la
materia, de acuerdo con la hip\'otesis de un espacio-tiempo discreto. Un conjunto de l\'amparas
luminosas est\'an conectadas entre s\'{\i} formando una estructura c\'ubica. Utilizando procedimientos
digitales de encendido y apagado de las l\'amparas se puede obtener se\~nales luminosas que se
propagan por el ret\'{\i}culo y que pod\'{\i}a interpretarse como la funci\'on de onda que utiliza la
mec\'anica cu\'antica para la descripci\'on de un sistema elemental.

Tambi\'en Alberto Galindo se interes\'o por la comunicaci\'on haciendo dos preguntas:

\medskip
1) ?`Se han de dar en el ret\'{\i}culo, antes de tomar el l\'{\i}mite continuo, las propiedades de simetr\'{\i}a
y leyes de conservaci\'on que se demuestran en el modelo continuo de las leyes f\'{\i}sicas?
El autor respondi\'o que parece razonable que tambi\'en en el modelo discreto se den unas
simetr\'{\i}as an\'alogas, y se refiri\'o a los trabajos de investigaci\'on que est\'a realizando
sobre subgrupos discretos de los grupos de Lie.

\medskip
2) ?`Qu\'e papel pueden jugar en estos espacios discretos las teor\'{\i}as recientes de geometr\'{\i}a no
conmutativa? El autor respondi\'o que se est\'an publicando recientemente hip\'otesis f\'{\i}sicas donde
las coordenadas espacio-temporales no conmutan entre s\'{\i} (cfr. A. Connes, ``Geometrie
non-commutative'', Intereditions, Paris 1990). En esta obra Connes propone la idea de un espacio
no conmutativo con el objeto de suprimir las divergencias ultravioletas introduciendo un corte
(cut-off) natural. Pero esta hip\'otesis lleva a la deformaci\'on de los grupos de simetr\'{\i}a
por los grupos cu\'anticos. Precisamente las realizaciones de los grupos cu\'anticos para
construir las ecuaciones de onda llevan a introducir de una manera natural los operadores
diferencias finitas equivalentes a las que el autor ha empleado en sus modelos discretos.

\end{document}